\documentclass{appolb}
\usepackage[dvips]{color}
\usepackage{ulem}
\usepackage{epsfig,epsf,subfig,graphicx,amsfonts,amssymb,color, subfig, color}
\usepackage{hyperref}

\begin{document}

\title{Route to chaos in generalized logistic map}

\author{R.~Rak$^{1}$, E.~Rak$^{1}$
\address{$^1$ Faculty of Mathematics and Natural Sciences, University of Rzesz\'ow, PL--35-959 Rzesz\'ow, Poland}}

\maketitle
\begin{abstract}
Motivated by a possibility to optimize modelling of the population
evolution we postulate a generalization of the well-know logistic
map. Generalized difference equation reads:
\begin{equation}
x_{n+1}=rx^p_n(1-x^q_n),
\end{equation}
$x\in[0,1],\;(p,q)>0,\;n=0,1,2,...$, where the two new parameters
$p$ and $q$ may assume any positive values. The standard logistic
map thus corresponds to the case $p=q=1$. For such a generalized
equation we illustrate the character of the transition from
regularity to chaos as a function of $r$ for the whole spectrum of
$p$ and $q$ parameters.
As an example we consider the case for $p=1$ and $q=2$ both in the periodic and chaotic regime. We focus on the character of the
corresponding bifurcation sequence and on the quantitative nature of the resulting attractor
as well as its universal attribute (Feigenbaum constant).
\end{abstract}


\section{Introduction}
Since the dawn time the chaos is an indispensable part of human life. One of the most famous models with chaotic behavior is the logistic map.
Its simple analytic form meant that it was used in many scientific disciplines such as biology, cryptography, communication chemical physics and stock market ~\cite{pareeka, ferrettia, pellicer, andrecut}.
One of the mathematical models of chaos was discovered by Mitchell
Feigenbaum (1978)~\cite{Fei1, Fei2, Thom1, Steph1}. Feigenbaum considered ordinary difference
equations used for example in biology to describe the development
of population in its dependence on time. He discovered that
population oscillates in time between stable powers (fixed
points), the number of which doubles according to changes in the
power of external parameter. The generalization of the Feigenbaum
model contains all first-order difference equations
$f(x_n)=x_{n+1}$. The condition for the existence of chaos is the
single maximum of a function $f(x_n)$. Feigenbaum also proved that
the transition to chaos is described by two universal constans
$\alpha$ and $\delta$, later named the Feigenbaum numbers.\\
We quote here a few definitions which are basic in the
development of this exposition.

\bf Definition 1.\rm  ~[cf. \cite{Smital}, p. 62] \\
Let $I=[a,b]\subset\Re$, $a<b$, $f(I)\subset I$. Discrete
iterations of a function $f:I\rightarrow I$ are called functions
$f^n:I\rightarrow I$ defined inductively: $f^0=id$,
$f^{n+1}=f\circ {f^n}$,
$n=1,2,3,...$.

\bf Definition 2.\rm  ~[\cite{Smital}, p. 62] \\
The sequence $\{x_n\}$, where $x_n=f^n(x_0)$ is called the
sequence of iterates of a function $f$ generated by point
$x_0\in I$.

\bf Definition 3.\rm  ~[cf. \cite{Smital}, p. 65]\\
A point $s\in I$, with $f(s)=s$ is called a first-order fixed
point of $f$. A fixed point $s$ of a function $f$ is locally
stable if there exists a neighbourhood $U_s$ such that all
sequences of iterates are convergent to $s$ for $x_0\in U_s$.
\\
$\mathbf{Theorem 1.}$ [see \cite{Smital}, p. 68] \\
If function $f$ is differentiable in fixed point $s$ and $|f^{'}(s)|<1$, then $s$ is attractive.

\bf Definition 4.\rm  ~[cf. \cite{Smital}, p. 72]\\
Let $f:I\rightarrow I$. We say that $s$ is a periodic point of the
function $f$, or that $s$ generates a cycle of order $k\geq 2$, if
$f^k(s)=s$ and $\forall_{i<k} f^{(i)}(s)\not =s$.
\\
$\mathbf{Theorem 2.}$ \cite{Li1}\\
If function $f$ has a cycle of order different to $2^k$, then  $f$ is chaotic.
\\
$\mathbf{Theorem 3.}$  [cf. \cite{Smital}, p. 79]\\
For arbitrary continuous function $f:I\rightarrow I$ existence cycles of order $m$
involve the existence cycles of order $n$ in the following
order:\\
$3>>5>>7>>2\cdot 3>>2\cdot 5>>2\cdot 7...>>2^k\cdot 3>>2^k\cdot
7>>...>>2^k>>...>>8>>4>>2>>1$.

\begin{center}
{\it  \large {The conventional logistic map}}
\end{center}
The Logistic map ~\cite{verhulst, Ott1, Schu1},
\begin{equation}
f(x_n)=x_{n+1}=rx_n(1-x_n),\; x\in[0,1],\; (p,q)>0,\; n=0,1,2,...
\end{equation}
is one of the most simple forms of a chaotic process. Basically,
this map, like any one-dimensional map, is a rule for getting a
number from a number. The parameter $r$ is fixed, but if one
studies the map for different values of $r$ (up to 4, else the
unit interval is no longer invariant) it is found that $r$ is the
catalyst for chaos. After many iterations $x$ reaches some values
independent of its starting value 3 regimes:
$r<1:\; x=0$ for large $n$,\quad $1<r<3:\; x=constant$ for large
$n$,\quad $3<r<r_{\infty}:\;$ cyclic behavior, where $r_{\infty}\approx3.56$,\quad
$r_{\infty}<r\leq 4:\;$ mostly chaotic,\quad \\
An interesting thing happens if a value of $r$ greater than 3 is
chosen. The map becomes unstable and we get a pitchfork
bifurcation with two stable orbits of period two corresponding to
the two stable fixed points of the second iteration of $f$. With
$r$ slightly bigger than 3.54, the population will oscillate
between 8 values, then 16, 32, etc. The lengths of the parameter
intervals which yield the same number of oscillations decrease
rapidly; the ratio between the lengths of two successive such
bifurcation intervals approaches the $Feigenbaum~~constant$
$\delta =4.669...$. The period doubling bifurcations come faster
and faster ( 8, 16, 32, ...), then suddenly break off. Beyond a
certain point, known as the accumulation point $r_{\infty}$,
periodicity gives way to chaos. In the middle of the complexity, a
window suddenly appears with a regular period like 3 or 7 as a
result of mode locking. The 3-period bifurcation occurs at
$r=1+2\sqrt{2}$, and period doubling then begin again
with cycles of 6,12,... and 7,14,28,... and then once again break
off to chaos.

\section{Generalized logistic map}
\quad One-dimensional representations are the simplest
arrangements capable to producing a chaotic movement.
Let us consider $p,q>0$. We postulate non-linear function:
\begin{equation}
f_r(x)=rx^p(1-x^q),\; x\in[0,1]
\label{glm}
\end{equation}
set by difference equation:
\begin{equation}
x_{n+1}=rx^p_n(1-x^q_n),\;x\in[0,1],\; n=0,1,2,....
\label{genlogmap}
\end{equation}
For the first time the function (\ref{glm}) has been presented in \cite{rak}.
The maximum of the function (\ref{glm}) is reached for
$x=\left(p \over p+q\right)^{1/q}$, because
$f^{\;'}_{r}(x)=r(px^{p-1}-(p+q)x^{p+q-1})$
and $f^{\;'}_{r}(x)=0\; \Leftrightarrow\; x=\left(p \over
p+q\right)^{1/q}$, $\Re\ni(p,q)>0$.\\
It results that $f_{r}\left(\left(p \over p+q\right)^{1\over q
}\right)$=$rqp^{p\over q}\over (p+q)^{p+q\over q}$ and for $r\in
\left(0,\; {({p+q\over p})^{p \over q}(p+q)\over q}\right)$
where\\
$\left(0,\; {({p+q\over p})^{p \over q}(p+q)\over
q}\right)=r_{max}$ function $f_r$ is $f([0,1])\subset [0,1]$.\\

\begin{figure}[!ht]

\begin{center}
\epsfxsize 12cm
\hspace*{0.3in}
\epsffile{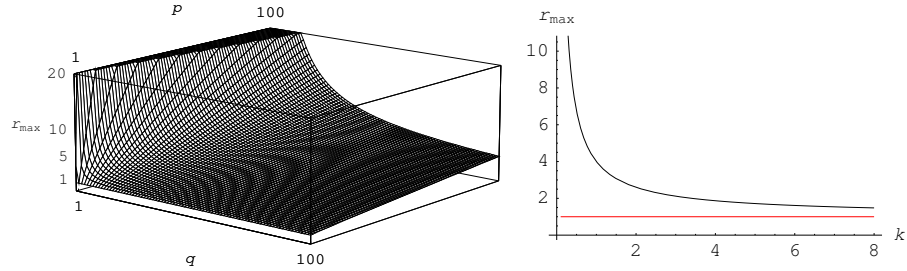}
\end{center}
\caption{Values of $r_{max}$ for generalized logistic map depending on $p$ and $q$ (left) and depending on $k$ (right).}
\label{fig1}
\end{figure}

Parameter $r_{max}$ depending on $p$ and $q$ accepts the following
limit values (Fig.\ref{fig1}):
$$\lim_{q\to \infty} r_{max}=1, \; \; \lim_{q\to
0^+}r_{max}=\infty$$
$$\lim_{p\to \infty}r_{max}=\infty, \; \;
\lim_{p\to 0^+}r_{max}=1$$

For the sake of parameters $p$ and $q$ the function (3) generates
a lot of new functions which can be analyzed using mathematical
methods, just like for a logistic map,
thus obtaining very interesting results.\\
a)\quad Let us assume that $k={q\over p}$ and $k>0$. Then $r$ is depend
on the value $k$ and $r\in\left[0,\frac{(k+1)^{\frac{k+1}{k}}}{k}\right].$\\
b) \quad If $k=1$, then $p=q$ and $r_{max}$ is independent
on $p$, $q$ and $r_{max}=4$ always - for logistic map, we have $p=q=1$ and $r_{max}=4$.\\
c) \quad If $\quad k \rightarrow {\infty}$, then $r_{max} \rightarrow 1$, because
$\lim_{k\to \infty } \,\frac{(k+1)^{\frac{k+1}{k}}}{k}=1.$\\
d) \quad If $k\rightarrow {0^{+}}$, then
$r_{max}\rightarrow {\infty}$, because
$\lim_{k\to \infty } \,\frac{(k+1)^{\frac{k+1}{k}}}{k}=+\infty.$\\
e) \quad Now let us consider a case, when $p=1$ and $q>0$. Then we have
\begin{equation}
f_{r}(x)=rx(1-x^q),~~x\in[0,1],~~ n=0,1,2,....
\end{equation}

and $r\in\left[0,\frac{(q+1)^{\frac{q+1}{q}}}{q}\right].$

\begin{figure}[!ht]
\epsfxsize 10cm
\hspace*{1in}
\epsffile{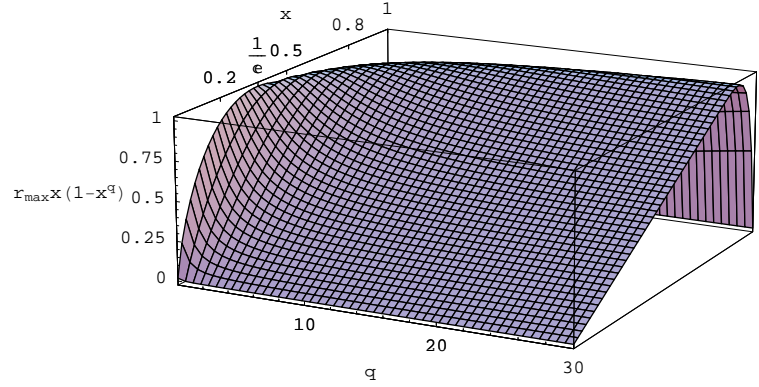}
\caption{$f_r(x)=rx(1-x^q)$ for $r=r_{max}$ depending on $q$.}
\label{fig2}
\end{figure}

Taking into account the above relationship we can observe change maximum of the function (5) depending on $q$ (Fig.\ref{fig2}).

\newpage

\bf A)\quad The property of $\bf x_{n+1}=rx^p_n(1-x^q_n)$ for $\bf p=1$, $\bf q=2$ in the periodic region.\rm\\

In the following part of this paper we will show that the
Feigenbaum model of transition to chaos is correct for
representation (4) for given positive values of parameters $p$
and $q$. Therefore, we will study the dynamics of this map in the
case of $p=1$ and $q=2$. Then the representation (5) assumes the
following form:
\begin{equation}
f_{r}(x)=rx(1-x^2),\;\;  x\in[0,1],\;\;  n=0,1,2,...
\label{eq12}
\end{equation}
A function $f_r$ in the point $x={\sqrt3\over 3}$ has a local maximum
and the value of function in this point is equal to
$f_{r}({\sqrt3\over 3})={2r\sqrt3\over9}$. From this relationship, we have
$f([0,1])\subset [0,1]$ for $r\in(0,{3\sqrt3\over 2}]$. The
trajectories behavior generated by $f_r$ depends on $r$ and,
therefore, we have to distinguish several intervals of this
value:\\
1)\quad For $r\in(0,1]$ all the sequences of iterates
$x_1,x_2,x_3,...,x_n$ depend on the initial value $x_0$ tend to
$s=0$. This point is the fixed point of the map for $r\in(0,1]$
and according to \it Definition 3 \rm it is also the stable point
of this map.\\
2)\quad Let $r=r_{1}>1$. If $x=rx(1-x^2)$ then, $x_{1}=0$,
$x_{2}=\sqrt{r-1\over r}$, so for $r>1$ function (6) has two fixed
points. Using \it Theorem 1 \rm we get: $f^{\; '}_{r}(x)=r-3rx^2$,
$|f^{\; '}_{r}(0)|=r$ and $|f^{\; '}_{r}(\sqrt{r-1\over
r})|=|3-2r|$.\\
Therefore, we have that $x_1=0$ is an unstable fixed point,
whereas $x_{2}=\sqrt{r-1\over r}$ is an attractive fixed point for
$r\in (1,2)$.\\
Change $r$ from 1 to 2 has caused the migration of the attractive
point from 0 to ${\sqrt2\over 2}$.\\
\it Conclusion 1. \rm \\
For $1<r<2$ there exists a stable fixed
point attracting all points. When $r=r_{1}=2$, then $|f^{\;
'}_{r}(\sqrt{r-1\over
r})|=|3-2r|=1$ and this point stops being an attractive point.\\
3)\quad For $r=r_{2}\geq 2$ we will look into the stability of
fixed points $f_r$ and $f_r^{(2)}$ of the map (5) as a function of
parameter $r$, because the Feigenbaum model is generated by
bifurcations connected with iterations of a function.

Let us consider the second iteration of $f_r$, namely
$f^{(2)}_{r}(x)=r^2x - r^2x^3 - r^4x^3 + 3r^4x^5 - 3r^4x^7 + r^4x^9$, with fixed points: $x_{1}=0$, $x_{2}=\sqrt{r-1\over r}$,
$x_{3}=\sqrt {{1\over 2} - {\sqrt {r^2-4}\over 2r}}$, $x_{4}=\sqrt
{{1\over 2} + {\sqrt {r^2-4}\over 2r}}$.
Because $f^{\; '(2)}_{r}(x)=r^2 - 3r^2x^2 - 3r^4x^2 + 15r^4x^4 -21r^4x^6 + 9r^4x^8$
we have $f^{\; '(2)}_{r}(x_1)=r^2$, $f^{\;'(2)}_{r}(x_2)=(3-2r)^2$,  $f^{\; '(2)}_{r}(x_3)=f^{\;'(2)}_{r}(x_4)=9-2r^2$ whereas for $2<r_2<\sqrt 5$ fixed points
$x_3$ and $x_4$ are attractive points according to condition in
\it Theorem 1 \rm.\\
\it Conclusion 2.\rm \\
The fixed point $x=\sqrt{r-1\over r}$ of $f_r^{(1)}$ is also the
fixed point of $f_r^{(2)}$ as well as of all higher iterations.\\
\it Conclusion 3.\rm \\
If the fixed point of $f_r^{(1)}$ becomes unstable, then it is
also the unstable fixed point of $f_r^{(2)}$ and of all next
iterations.\\
From the inequality $|f^{\; '}(s)|>1$ we have
$|f^{\; '(2)}(s)|=|f^{\; '}[f(s)]f^{\; '}(s)|=|f^{\; '}(s)|>1$.\\
If $\sqrt 5<r_3<r_4$, then the fixed points of $f^{(2)}_r$ become
repulsive at the same time. Following this instability, the fourth
iteration shows two new pitchfork bifurcations giving cycle $2^2$
order for four attractive fixed points, which is called period
duplication.

\begin{figure}[!ht]
\epsfxsize 14cm
\hspace{2.4cm}
\epsffile{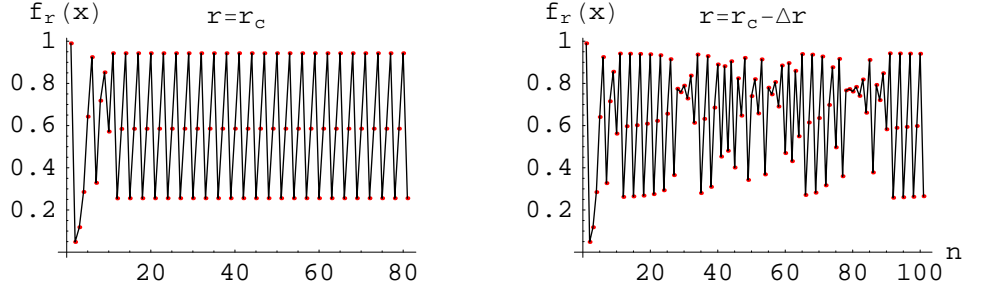}
\caption{Iterates of map (6) strting from $x_{0}=0.99$. Left: In the
stable 3-cycle region for $r=r_c$;~~Right: In the chaotic region for
$r=r_{c}-\Delta r,\; \Delta r=0.001$ .}
\label{fig3}
\end{figure}

To generalize the above examples we get:\\
\bf a)\rm \quad For $r_{n-1}<r<r_n$ there exists a stable cycle
$2^{n-1}$, whose elements
$\overline{x}_0,\overline{x}_1,\overline{x}_2,\overline{x}_3,...,\overline{x}_{2^{n-1}-1}$
 are defined by the following
connections:\\
$$f_{r}(\overline{x} _i)=\overline{x}_{i+1},\quad \quad
f_{r}^{2^{n-1}}(\overline{x} _i)=\overline{x}_{i},\quad \quad
|\prod_{i}f_{r}^{\; '}(\overline{x}_i)|<1$$\\
\bf b)\rm \quad For $r_n$ all points of the cycle of $2^{n-1}$
become unstable simultaneously and the bifurcation creates a new
stable cycle of $2^n$ order for
$f_{r}^{2^n}=f_{r}^{2^{n-1}}(f_{r}^{2^{n-1}})$ existing for
$r_{n-1}<r<r_n$.\\
\bf c)\rm \quad For a map (6) the value of parameter $r_{n+1}$ we can
describe the following connection: $r_{n+1}\approx \sqrt{3+r_n},$
where $r_{0}=1,\; \; r_1=2,r_2=\sqrt{5},...,r_{\infty}$.\\
\bf d)\rm \quad There exists a point of accumulation of an infinite
number of the bifurcation of period duplication for the finite value
$r$ which we appointed as $r_{\infty}$: $r_{\infty}=\lim_{n\to
\infty}r_{n}={1+\sqrt{13}\over
2}\approx{2.303}$.\\
Taking into account the above, $r_{n+1}\approx r_{n}$,\quad  $\sqrt{3+r}=r$,
$r^{2}-r-3=0$,\quad thus $r={1+\sqrt{13}\over 2}$.\\
\bf e)\rm \quad For the representation (6) there exists a constant,
whose value corresponds to the Feigenbaum constant:
$$\delta= \lim_{n\to \infty}{r_{n}-r_{n-1}\over{r_{n+1}-r_{n}}}\approx{4.61}.$$\\
\it Proof.\rm \quad Let us assume that $r_{n-1}=a$. Then $r_{n}=\sqrt{3+r_{n-1}}=\sqrt{3+a}$,
$r_{n+1}=\sqrt{3+r_{n}}=\sqrt{3+\sqrt{3+a}}$ and
$\delta=\lim_{n\to\infty}{r_{n}-r_{n-1}\over{r_{n+1}-r_{n}}}=\sqrt{14+2\sqrt{13}}\approx{4.61}$.\\
A table of $2^n$ cycles type and values of $r_n$ is given below. It is clearly visible that doubling bifurcations come faster and
faster (8, 16, 32, ...) and suddenly break off beyond a certain point known as the accumulation point and periodicity gives way to chaos.
\begin{table}[!ht]
\epsfxsize 12.5cm \hspace{0cm} \epsffile{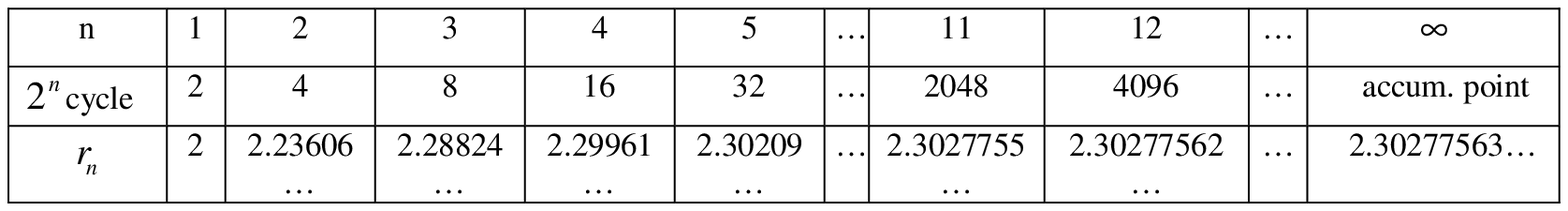} \caption{The
algebraic orders of the values of $r_n$ for $n=1,2,...$ are given
by $1,2,4,8,...$.} \label{tab1}
\end{table}
\\
\bf B)\quad The property of $\bf x_{n+1}=rx^p_n(1-x^q_n)$
for $\bf p=1$, $\bf q=2$ in chaotic region.\rm
\\
\bf a)\rm \quad For $r_{\infty}$ the bifurcation sequence ends
with a set of the infinitely numerous points, which is called the
set of Feigenbaum attraction. \\
\bf b)\rm \quad If $r_{\infty}<r\leq {3\sqrt{3}\over{2}}$ then we
observe the irregular dynamics of map (6) with narrow ranges in
which the set of attraction has a periodical character.\\
\bf c)\rm \quad Numeric calculations for the representation (6) show
that the largest range is for $r=r_{c}\approx 2.451$. Then
there exists 3-cycle (Fig.\ref{fig3}), which according to th.2 and th.3,
implicates the chaos as well as the existence of cycles of
different orders.\\
\bf d)\rm \quad For the map (6) at $r=r_{max}={3\sqrt 3\over 2}$,\\

\begin{figure}[!ht]
\epsfxsize 14cm
\hspace{2.5cm}
\epsffile{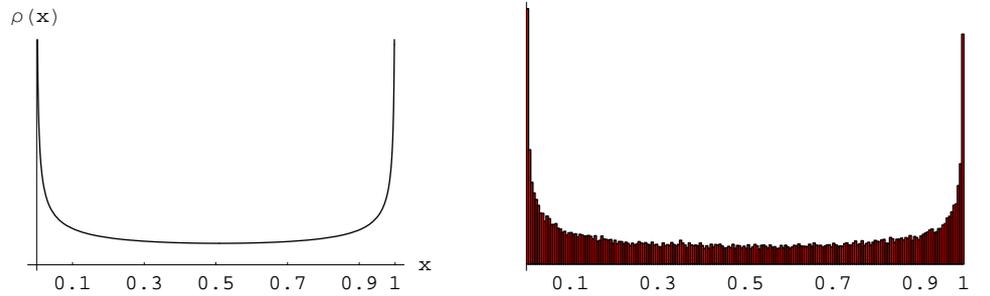}
\caption{The invariant density function of map (6) for $r=r_{max}={3\sqrt 3\over 2}$. Left: An analytical form represented by the equation (10). Right: Numerical simulation - the histogram for 50000 iterations.}
\label{fig4}
\end{figure}

\begin{equation}
x_{n+1}={3\sqrt 3\over 2}x_{n}(1-x_{n}^2)\equiv
f_{r_{max}}(x_{n}),
\end{equation}
can be solved by the simple change of
variables:
\begin{equation}
x_{n}=\cos{(y_{n})}\equiv g(y_{n}).
\end{equation}
Then (7) can be converted into\\
$$\cos{(y_{n+1})}={3\sqrt 3\over 2}\cos{(y_n)}[1-\cos{^2(y_n)}]={3\sqrt 3\over
2}\cos{(y_n)}\sin{^2(y_n)},$$\\
which has one solution $y_{n+1}=\arccos{\left[{3\sqrt 3\over
2}\cos{(y_n)}\sin{^2(y_n)}\right]}$ and \\
$$x_{n}=\arccos (\frac{1}{{\sqrt{3}}\,{\left( -{y_0} +{\sqrt{-1 + {{y_0}}^2}} \right) }^{\frac{1}{3}}}+
\frac{{\left( -{y_0} + {\sqrt{-1 + {{y_0}}^2}} \right) }^
{\frac{1}{3}}}{{\sqrt{3}}}).$$\\
The invariant density function $\rho_{{3\sqrt 3\over 2}}(x)$ of $f_{{3\sqrt
3\over 2}}(x)$ can be calculated using the following definition:\\
\begin{equation}
\lim_{N\to \infty}{1\over N}\sum_{n = 0}^{N-1}
     \delta \,\left( x - {x_n} \right)=\lim_{N\to \infty}{1\over N}\sum_{n = 0}^{N-1}
     \delta \,\left( x - {g(y_n)} \right)
\end{equation} Using $\rho(y)=1$, (9) becomes: $\rho_{{3\sqrt 3\over 2}}(x)=\int_0^{1}dy\rho(y)\delta[x-h(y)]$\\
i.e.
\begin{equation}
\rho_{{3\sqrt 3\over 2}}(x)=\frac{\left(\sqrt{x^2-1}-x\right)^{2/3}-1}{3 \sqrt{x^2-1}
   \sqrt[3]{\sqrt{x^2-1}-x}
   \sqrt{-\left(\sqrt{x^2-1}-x\right)^{2/3}-\frac{1}{\left(\sqrt{x^2-1}-x
   \right)^{2/3}}+1}} .
\end{equation}
The above function is shown in Fig.\ref{fig4}.
\\

The invariant density function of the logistic map for $r=r_{max}=4$ is given by $\frac{1}{\pi  \sqrt{(1-x) x}}$.
\\
\\

\begin{figure}[!ht]
\epsfxsize 13.5cm
\hspace{-0.3cm}
\epsffile{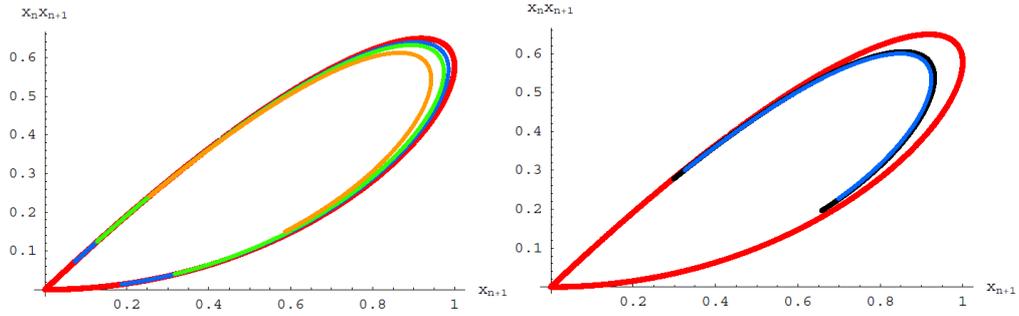}
\caption{The set of attraction for representation (6) [60000 iterations]: a) for $r=r_{max}$(red), $r=2.56$(blue), $r=2.53$(green), $r=2.45$(orange);~~b) for $r=r_{max}$(red), $r=2.42$(black), $r=2.4$(blue).}
\label{fig5}
\end{figure}

\bf C)\quad Ordered chaos - construction of an attractor for\\
$\bf x_{n+1}=rx_{n}(1-x_{n}^2)$ and $\bf r=r_{max}={3\sqrt
3\over2}$.\rm \\
\\
At the end of this considerations we will present a method of the
{\it ordering of chaos}. A lot of numeric experiments have led
to the situation where it is possible to rank chaos in such a way
that the successive iterations generated by chaotic dynamic system
would reach an attractor of a regular shape. This method also has
it, that from the sequence of several thousand iterations (for
representation $x_{n+1}=rx_{n}(1-x_{n}^2)$ where $r=r_{max}={3\sqrt
3\over 2}$ defined recursively) we draw up the following points
$(x_{n+1},x_{n}x_{n+1})$ on a plane - we construct an
attractor. In this way an attractor comes into being, which is visited irregularly by points of successive iterations (Fig.\ref{fig5}).\\

\newpage

\bf D)\quad Bifurcation diagram and Lyapunov exponent.\rm \\
\\

\begin{figure}[!ht]
\epsfxsize 9cm
\hspace{2cm}
\epsffile{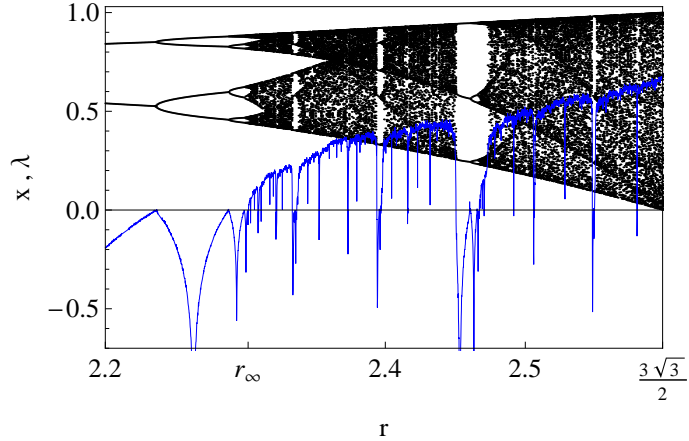}
\caption{The bifurcation diagram and the Lyapunov exponent for the map ~(\ref{eq12}) and $2.2<r\leq {3\sqrt{3}\over{2}}$.}
\label{fig6}
\end{figure}

Nowadays the Lyapunov exponent is applicable in various scientific disciplines such as biology, engineering, bioengineering and informatics ~\cite{lap11, lap1, lap2, lap3, lap4, lap5}.
A quantitative measure of the sensitive dependence to initial conditions is the Lyapunov exponent, which is a measure of the exponential separation of nearby orbits. For discrete time system $x_{n+1}=f(x_n)$, for an orbit starting with $x_0$ the Lyapunov exponent $\lambda(x_0)$ can be calculated by
\begin{equation}
\lambda(x_0)=\lim_{n\to \infty} \frac{1}{n}\sum_{i = 0}^{n-1}\ln|f'(x_i)|,
\label{lyapunov}
\end{equation}
A positive Lyapunov exponent can be considered as an indicator of chaos, whereas negative exponents are associated with regular and periodic behavior of the system.
Figure \ref{fig6} shows the bifurcation diagram and Lyapunov exponent for the equation (\ref{eq12}). It is clear that even in chaotic areas ($r_{\infty}<r\leq {3\sqrt{3}\over{2}}$) there are many periodic intervals ($\lambda<0$).

\section{Summary}
Currently, modeling real-world systems is very popular. A well-known model that takes into account the deterministic chaos is the logistic map.
However, this one-dimensional map can be controlled only by one parameter. In the present contribution, we postulate a generalization of the classical logistic map which is generally controlled by three parameters. This could allow to adapt to the conditions prevailing in the modeled system.

We focus on the $r_{max}$ parameter (depending on $p$ and $q$) which is responsible for the dynamics of the system - among others shown that if $p=q$ then $r_{max}=4$.

As an example, we carry out analytical and quantitative analysis in the case where $p=1$ and $q=2$ both in the periodic and chaotic regime where the Lyapunov exponent has positive values. It turns out that the dynamics of this equation is faster than in the case of the logistic map - the value of the parameter $r_{max}\approx2.3$ is less than the value of $r_{max}=4$ for the logistic map.
In the periodic regime we identified Feigenbaum constant $\delta$ which is typical for all dissipative systems (also for the logistic map).
For chaotic area we have found an analytical form of the invariant density function. For both regimes we have proposed specific form of data representation which allowed to obtain a non-trivial structure of an attractor set.\\

\bf Footnotes \rm \\
This work was partially supported by the Centre for Innovation and Transfer of Natural Sciences and Engineering Knowledge (University of Rzeszow).

\end{document}